\documentclass[%
 aip,
 amsmath,amssymb,
 reprint,%
]{revtex4-1}

\usepackage{graphicx}
\usepackage{dcolumn}
\usepackage{bm}

\usepackage[utf8]{inputenc}
\usepackage[T1]{fontenc}
\usepackage{mathptmx}

\begin{document}

\preprint{AIP/123-QED}

\title[]{Fitting the trajectories of particles in the equatorial plane of a magnetic dipole with epicycloids}

\author{Fancong Zeng}
 \email{ Fczeng$\_$lut@163.com}
 \affiliation{Lanzhou University of Technology, Lanzhou 730050, China}

\author{Konstantin Kabin}%
 \email{ Konstantin.Kabin@rmc.ca}
\affiliation{ Department of physics and space science, Royal Military College of Canada, Kingston, Ontario, K7K 7B4, Canada
}%

\author{Xiao-Yun  Wang}
\email{ xywang@lut.edu.cn (Corresponding author)}
\affiliation{Lanzhou University of Technology, Lanzhou 730050, China}
\affiliation{ Department of physics, Lanzhou University of Technology, Lanzhou 730050, China
}%

\author{Dao-bin Wang}
\email{photonics_wang@yahoo.com}
\affiliation{Lanzhou University of Technology, Lanzhou 730050, China}
\affiliation{ Department of physics, Lanzhou University of Technology, Lanzhou 730050, China}

\date{\today}

\begin{abstract}
In this paper we discuss epicycloid approximation of the trajectories of charged particles in axisymmetric magnetic fields. Epicycloid trajectories are natural in the Guiding Center approximation and we study in detail the errors arising in this approach. We also discuss how using exact results for particle motion in the equatorial plane of a magnetic dipole the accuracy of this approximation can be significantly extended. We also show that the epicycloids approximate the trajectory of a charged particle more accurately than the position of the particle along the trajectory.
\end{abstract}

\maketitle

\section{Introduction}

Motion of charged particles in geomagnetic field has important applications in plasma and space physics, for example, to dynamics of radiation belts and ring current \cite{Lyons84,Gombosi}. Most commonly, this motion is described using the Guiding Center (GC) approximation \cite{northrop}, which is one of the most successful approximations in plasma physics. Under the GC approximation, the motion of a charged particle is split into two motions considered separately: fast gyration of the particle around the guiding center and a much slower drift of the guiding center. The accuracy of this approximation decreases as the energy of the particle increases and while it is excellent for ring current electrons, it is hardly applicable to cosmic rays.

Motion of electrons and ions in the magnetosphere is, of course, three dimensional. However, many important insights into physics of various populations
of the charged particles and their dynamics can be gained from considering particles in the equatorial plane of the terrestrial dipole \cite{li98,sarris2005,kabin2017}. This is often done to simplify the analysis. Although seldom used in practice, there is an analytical solution for Lorentz equations of motion of charged particles in the equatorial plane a magnetic dipole \cite{kusaka,Juarez1949,Avrett1962}. Naturally, for small particle energies this solution approaches the GC expectations, but for higher energies can deviate from it very significantly \cite{Avrett1962}.

In a typical application of GC, the particle is considered to be ``collapsed" into the guiding center and only the motion of the guiding center itself is discussed. The gyration speed of the particle in the approach only appears through the first adiabatic invariant (the magnetic moment), which is assumed to be conserved exactly \cite{Kabin2019}. Thus, all information about the phase of the particle as it gyrates around the gyrocenter is lost. For some applications, however, it may be desirable to keep the phase information even in the GC approximation. For example, results of a particle interaction with high-frequency waves may depend on the relative phases of the particle and the wave. Some numerical codes may switch between GC approximation and integration of full Lorentz equations depending on the magnetic field parameters. In such cases, the phase of the particle transitioning from the region of GC approximation to full Lorentz integration is often assigned randomly.

Full motion of a charged particle in the GC approximation is represented by an epicycloid \cite{johnston}. In this paper we study how well the particle trajectory can be approximated by
an epicycloid curve. In particular, we use the results of the analytical solution for the particle motion in the equatorial plane of a dipole field \cite{Avrett1962} to determine the parameters of the epicycloid. We also study the accuracy of the epicycloid approximation based on GC parameters. In this paper we consider two versions of the GC approximation: the simplest possible implementation based on the initial position of the particle, and a more accurate one in which parameters at the guiding center are used. We show that with well-chosen parameters, the epicycloid approximation of the particle trajectory remains accurate for surprisingly large energies, even for those approaching the threshold of confinement in the dipole field \cite{Kabin2018}. Such high energies are usually considered to be well outside the area of applicability of the GC approximation.

\section{Equatorial plane of a magnetic dipole}
\label{ana_dip}

Here we consider magnetic field aligned with the $z$ axis of a right-handed Cartesian coordinate system $(x,y,z)$:
\begin{equation}
{B}_{z}=\frac{\cal M}{\left(x^{2}+y^{2}\right)^{3 / 2}}, \label{eq_B}
\end{equation}
where ${\cal M}$ is the strength of magnetic dipole moment expressed in units of T$\cdot$ m$^3$. For $z=0$
 this magnetic field coincides with that in the equatorial plane of a magnetic dipole.
Motion of charged particles in this field is described by Lorentz equations:
\begin{equation}
m{\ddot{\bf r}}=e {\bf v}\times {\bf B} \label{eq_Lor}
\end{equation}
where $m$ is the mass of the particle, $e$ is its charge, ${\bf v}$ its velocity and ${\bf r}$ its position. For magnetic field (\ref{eq_B})  motion along the $z$ axis is decoupled from the motion perpendicular to it, so for the rest of the paper we  only consider the trajectories in the $(x,y)$ plane. It is convenient to introduce
\begin{equation}
A =\frac{e {\cal M}}{ m}.
\label{eqA}
\end{equation}
to simplify the notation.

For magnetic field (\ref{eq_B}) there is a class of very simple solutions of (\ref{eq_Lor}) which are circles with the center at the origin. If $r$ is the radius of the circle, the corresponding circular speed is
\begin{equation}
v_c=\frac{|A|}{r^2}. \label{v_circ}
\end{equation}
This circular speed provides a convenient scale for the velocity in this problem.

Motion of charged particles in the equatorial plane of a magnetic dipole has many practical applications and has been studied extensively in the past \cite{kusaka, Juarez1949, Avrett1962}. This problem has an analytical solution which can be expressed in term of elliptic integrals. Typical particle trajectories are confined to an annulus defined by $r_{\min}$ and $r_{\max}$ which can be calculated as:
\begin{equation}
r_{\max }=\frac{2 r_{*}}{1+\sqrt{1-4v /v_*}} \ \ \
{\rm and } \ \ \ \
r_{\min}=\frac{2 r_{*}}{1+\sqrt{1+4v/ v_*}}. \label{r_m}
\end{equation}
Here $v$ is the speed of the particle (it remains constant during the motion),  $r_*$ is the radius where the particle velocity is pointing in the radial direction, and $v_*$ is the speed of the circular motion given by (\ref{v_circ}) at distance $r_*$ \cite{Avrett1962}.
If $v$ exceeds $v_*/4$ then $r_{\max}$ becomes undefined; in this case the particle trajectory is unbounded and eventually the particle moves to infinity.
Here we only consider bounded trajectories. Transition from bounded to unbounded trajectories for various axisymmetric fields is discussed in detail by Ref. \onlinecite{Kabin2018}.

Period of a single cycle (gyroperiod) can be calculated as \cite{Avrett1962}:
\begin{equation}
T=
\frac{2 r_*}{v} \int_{r_*/r_{\max }}^{r_*/r_{ \min} } \frac{d\xi}{\xi^2\sqrt{1-(v_*/v)^2 \xi^2(1-\xi)^2 } }=
\frac{2 r_*}{v} I_2( v/v_*) \label{gyro}
\end{equation}
where integral $I_2$ is a function of $v/v_*$ only; the subscript 2 is used for consistency with Ref. \onlinecite{Avrett1962}. The integration variable $\xi$ in (\ref{gyro}) is $r_*/r$. Note, that equation (\ref{gyro}) gives the ``synodic" period, i.e. the time it takes the particle initially, for example, at $r_{\min}$ reach the $r_{\min}$ position again.

The average drift around the origin of the coordinate system can be calculated as
\begin{equation}
{\Omega_D}=-{\rm sgn}(A)\frac{v_*}{r_*} \frac{I_1(v/v_*)}{I_2(v/v_*)} \label{drift}
\end{equation}
with
\begin{equation}
I_1(v/v_*)= \int_{r_*/r_{\max }}^{r_*/r_{ \min} } \frac{(1-\xi) d\xi}{ \sqrt{1-(v_*/v)^2 \xi^2(1-\xi)^2 } }.\label{drift2}
\end{equation}
Note, that $\Omega_D$ is negative for $A>0$, i.e. positively charged particles in upward magnetic field drift clockwise.
And integral $I_1$ is also a function of $v/v_*$ only. Note that both integrals $I_1$ and $I_2$ have integrable singularities at both ends of the interval, thus certain care needs to be taken for accurate numerical evaluation.

A typical trajectory of a particle in the equatorial magnetic field is shown with red lines in figures \ref{comparison} and \ref{traj_def}.

\section{Approximating particle trajectories with epicycloids}
\label{sec_epi}

Cycloidal motion is, of course, expected in the GC approximation. In this section we seek to approximate the trajectory of a particle making the best use of the analytical parameters of its trajectory discussed in section \ref{ana_dip}.
 Specifically, we place the center of the cycloid at $(r_{\min}+r_{\max})/2$ and ensure that the amplitude of circular motion is $(r_{\min}-r_{\max})/2$. We also use the exact expressions for the drift and gyration frequencies. Then the equation for the epicycloidal approximation to the particle trajectory can be written as
\begin{small}
\begin{equation}
\left(\begin{array}{l}
   x \\
 y
\end{array}\right)=\frac{(r_{\max}+r_{\min})}{2}
\left(\begin{array}{l}
   \cos (\Omega_D (t-t_0)) \\
  \sin (\Omega_D (t-t_0))
\end{array}\right)-\frac{(r_{\max}-r_{\min})}{2}
\left(\begin{array}{l}
   \cos (\omega (t-t_0)) \\
  \sin (\omega (t-t_0) )
\end{array}\right)
\label{eq1}
\end{equation}
\end{small}
where $t$ is time and $t_0$ is the initial time which defines the initial phase of the particle in its trajectory. If $t_0=0$ then the particle is initially $x=r_{\min}$ and $y=0$, which is a convenient initial condition.
With this choice of initial conditions, the epicycloid touches the exact trajectory at $r_{\min}$ and $r_{\max}$ points. Note, that in equation (\ref{eq1}) $\omega$ is the   ``sidereal" angular frequency of the particle which is computed as
\begin{equation}
\omega=\Omega_D-{\rm sign}(A)2\pi/T.
\end{equation}
The function ${\rm sign}(A)$ in this expression accounts for the direction of gyration of the particle which depends on the sign of the charge and the direction of the magnetic field.

\begin{figure}[tbph]
  \centering
  \includegraphics[scale=0.35]{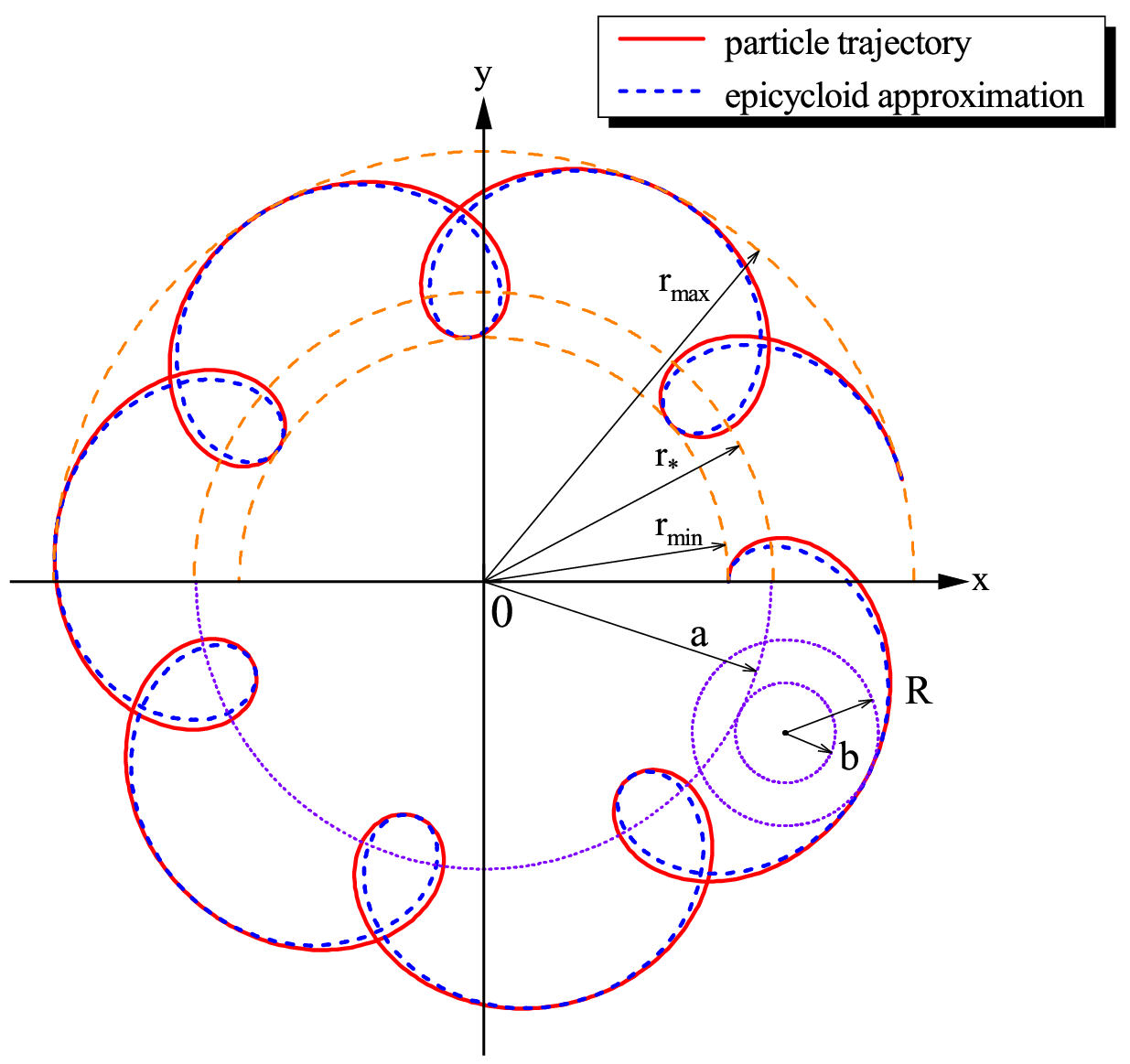}
  \caption{Comparison of the particle trajectory and the  epicycloid approximation for $A=10$ and $v=0.22 v_*$. Also shown are
the trajectory parameters $r_{\min}$, $r_{\max}$, and $r_*$ as well as  the cycloid parameters $a$, $b$, and $R$.}
\label{traj_def}
\end{figure}

Figure \ref{traj_def} shows an example of a comparison between the particle trajectory calculated numerically by integrating equations (\ref{eq_Lor}) with that approximated by an epicycloid. In this figure, $A=10$ and $v/v_*=0.22$. Red line shows the numerically calculated particle trajectory and blue dashed line is the epicycloid approximation.

\begin{figure}[tbph]
\flushleft
  \includegraphics[scale=0.32]{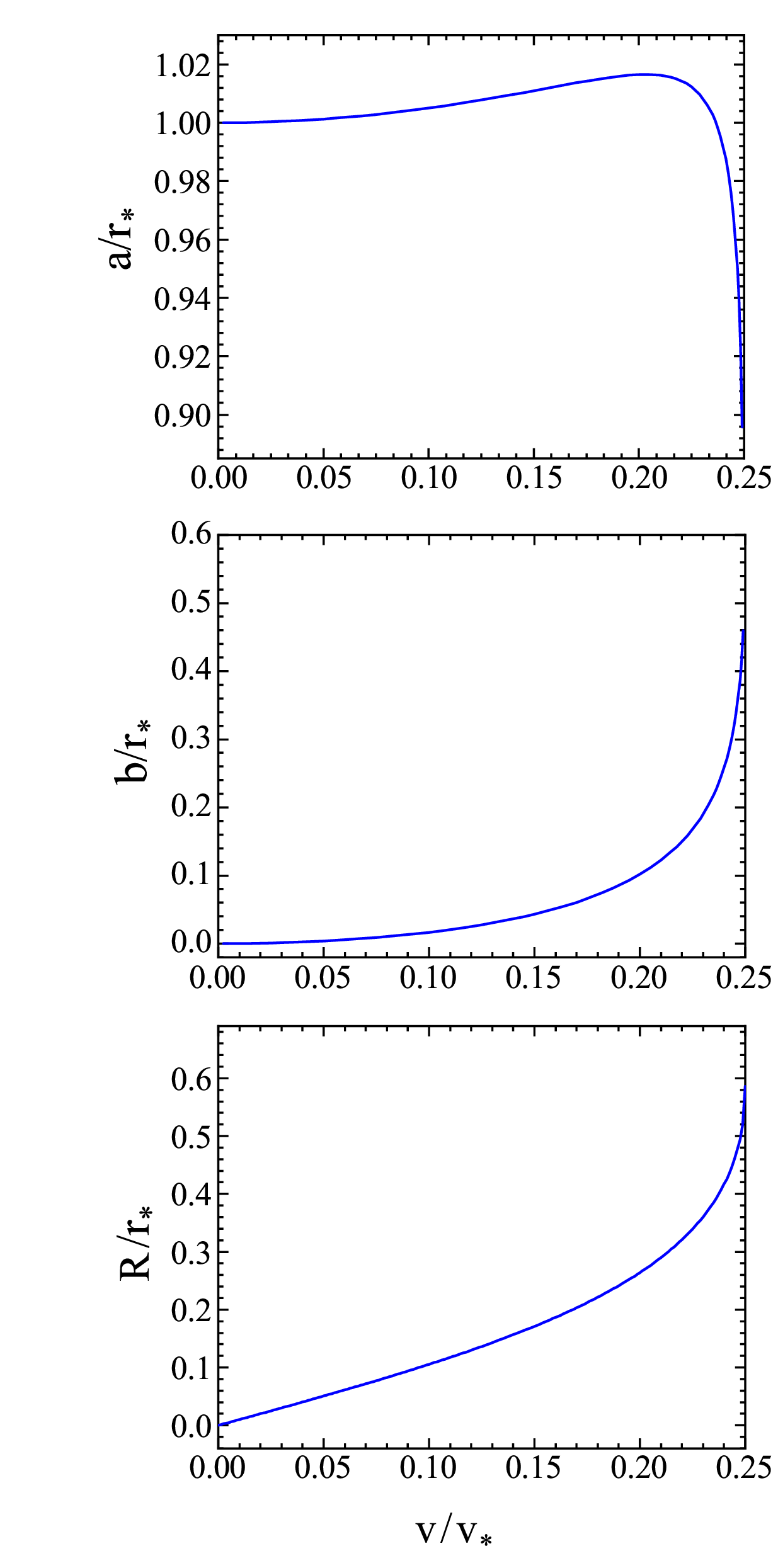}
  \caption{The the epicycloid parameters   $ a/r_*$, $b/r_*$, and $R/r_*$ as a function of $v/v_*$.  }
\label{abR}
\end{figure}

A traditional geometric interpretation of equations (\ref{eq1}) is a path traced by a point attached to a circle of radius $b$ rolling without slipping around the outside of a fixed circle of radius $a$, where the point is at a distance $R$ from the center of the moving circle  \cite{johnston}.
Parameters $a$, $b$, and $R$ can be calculated as follows:
\begin{equation}
\begin{array}{l}
R=(r_{\max }-r_{\min })/{2} \\
a+b=(r_{\max }+r_{\min })/{2} \\
{a}/{b}=-{2 \pi}/{( T \Omega_D )} .
\end{array}\label{par_abR}
\end{equation}
These geometric parameters are also shown in figure \ref{traj_def}. Note, that although $a$ is close to $r_*$, they are different, see the upper panel of figure \ref{abR}.
Figure \ref{abR} shows the variation of the epicycloid parameters, $a$, $b$, and $R$, normalized with $r_*$, with dimensionless speed of the particle $v/v_*$. It follows from equations (\ref{r_m}), (\ref{gyro}), and (\ref{drift})  that the curves in figure \ref{abR} for different initial positions of the particle coincide; parameters $a/r_*$, $b/r_*$, and $R/r_*$ depend only on $v/v_*$. Parameters $R$ and $a+b$ have obvious physical interpretations as the gyroradius of a particle and the distance of the guiding center from the origin, respectively. The physical meaning of the parameters $a$ and $b$ individually is unclear.

Note, that figure  \ref{traj_def} corresponds to a very large speed of the particle, which is nearly 90\% of the confinement threshold speed for a dipole. For a proton in
the Earth magnetic field with $r_*=3 R_E$ (typical value for the inner radiation belt) this speed would correspond to about 800 MeV, which is much larger than  typical energy of a radiation belt proton. Nevertheless, even at such extremely high energies, equation (\ref{eq1}) provides a fair approximation for the trajectory of a particle.
Furthermore since the exact frequencies are used, the epicycloid approximation does not contain secular error. Of course, in cases of practical interest, analytical expressions for such parameters are unlikely to be available. Nevertheless, it is remarkable how accurate cycloidal approximation (\ref{eq1}) with well-chosen parameters can be, even for very high particle energies which would normally be considered to be well outside the GC validity.

A more typical application of the cycloid approximation would be based on GC quantities, which we discuss next. Accuracy of the cycloid approximation is further discussed  in section \ref{accuracy}.

\section{Guiding center approximation}
\label{sec_GCa}

The gyrofrequency and the gyroradius for a particle moving in magnetic field (\ref{eq_B}) can be calculated as
\begin{equation}
\omega_{g}=-\frac{A}{r^3} \ \ \ \ {\rm and} \ \ \  r_g=\frac{v}{|\omega_g|} =\frac{v}{|A|}r^3 \label{om_g}.
\end{equation}
The gradient drift speed is e.g. Ref. \onlinecite{Lyons84}.
\begin{equation}
V_G=\frac{m}{q}\frac{v^2}{2B}\frac{{\bf B}\times \nabla B}{B^2}=-\frac{3 v^2}{2A} r^2 \label{V_G}
\end{equation}
giving the drift frequency of
\begin{equation}
\Omega_{GD}=-\frac{3}{2}\frac{v^2}{A}r \label{Om_G}.
\end{equation}
Note, that in the above expressions the gyrofrequency and the drift frequency can be either positive or negative, depending on the sign of $A$. These definitions  simplify the expression for the epicycloid approximation for the trajectory below.

An important question arising in applying equations (\ref{om_g}-\ref{Om_G}) to describe the motion of a specific particle is what value for $r$ should be used in these equations \cite{burby2013}.
The simplest interpretation of $r$ in equations (\ref{om_g}) and (\ref{Om_G}) is to consider it to be the instantaneous location of the particle. Under this interpretation, however, the gyroperiod, gyroradius, and the drift frequency all change as the particle moves.  If, for example,
we assume that the particle is initially at $r_{\min}$ then, because the magnetic field decreases with $r$, as the particle moves, $\omega_g$ decreases while $r_g$ and
$\Omega_{GD}$ both increase.

It is well known that the accuracy of the GC approximation improves if some sort of an average value is used for the magnetic field when applying equations (\ref{om_g}) and (\ref{Om_G}) \cite{burby2013}.
One way to improve the accuracy of the GC approximation is by calculating the gyroperiod and gyroraidus of the particle based on the previous estimation of the gyrocenter  instead of the current position of the particle. For example, assuming that the particle is initially at $r_{\min}$ we can set up the following iteration for the gyrofrequency, gyroradius, and drift frequency:
\begin{small}
\begin{equation}
\begin{array}{l}
\omega_g^{(i)}=-\frac{A}{\left(r_{\min}+r_g^{(i-1)}\right)^3},  \ \
 r_g^{(i)}=\frac{v}{|\omega_g^{(i)}|},   \ \
\Omega_{GD}^{(i)}=-\frac{3}{2}\frac{v^2}{A}(r_{\min}+r_g^{(i)}) \label{iter}
\end{array}
\end{equation}
\end{small}
with $r_g^{(0)}=v r_{\min}^3/|A|$. Here, superscript $^{(i)}$ refers to the iteration number (not the power).
Our numerical calculations show that iteration (\ref{iter}) for magnetic field (\ref{eq_B}) converges for $v<0.204 v_*$, independent of the initial value of $r_{\min}$. Note, that the velocity corresponding to this convergence threshold is smaller than that corresponding to the confinement threshold.

Using expressions (\ref{iter}), epicycloid equation for the trajectory can be written as:

\begin{equation}
\left(\begin{array}{l}
   x \\
 y
\end{array}\right)=(r_{\min}+r_g^{(i)})
\left(\begin{array}{l}
   \cos (\Omega_{GD}^{(i)} t) \\
  \sin (\Omega_{GD}^{(i)} t)
\end{array}\right)-r_g^{(i)}
\left(\begin{array}{l}
   \cos (\omega_g^{(i)} t) \\
  \sin (\omega_g^{(i)} t )
\end{array}\right).
\label{eqG1}
\end{equation}
Here we use the same initial conditions as before: for $t=0$ the particle is at $x=r_{\min}$ and $y=0$.

\begin{figure}[tbph]
  \includegraphics[scale=0.32]{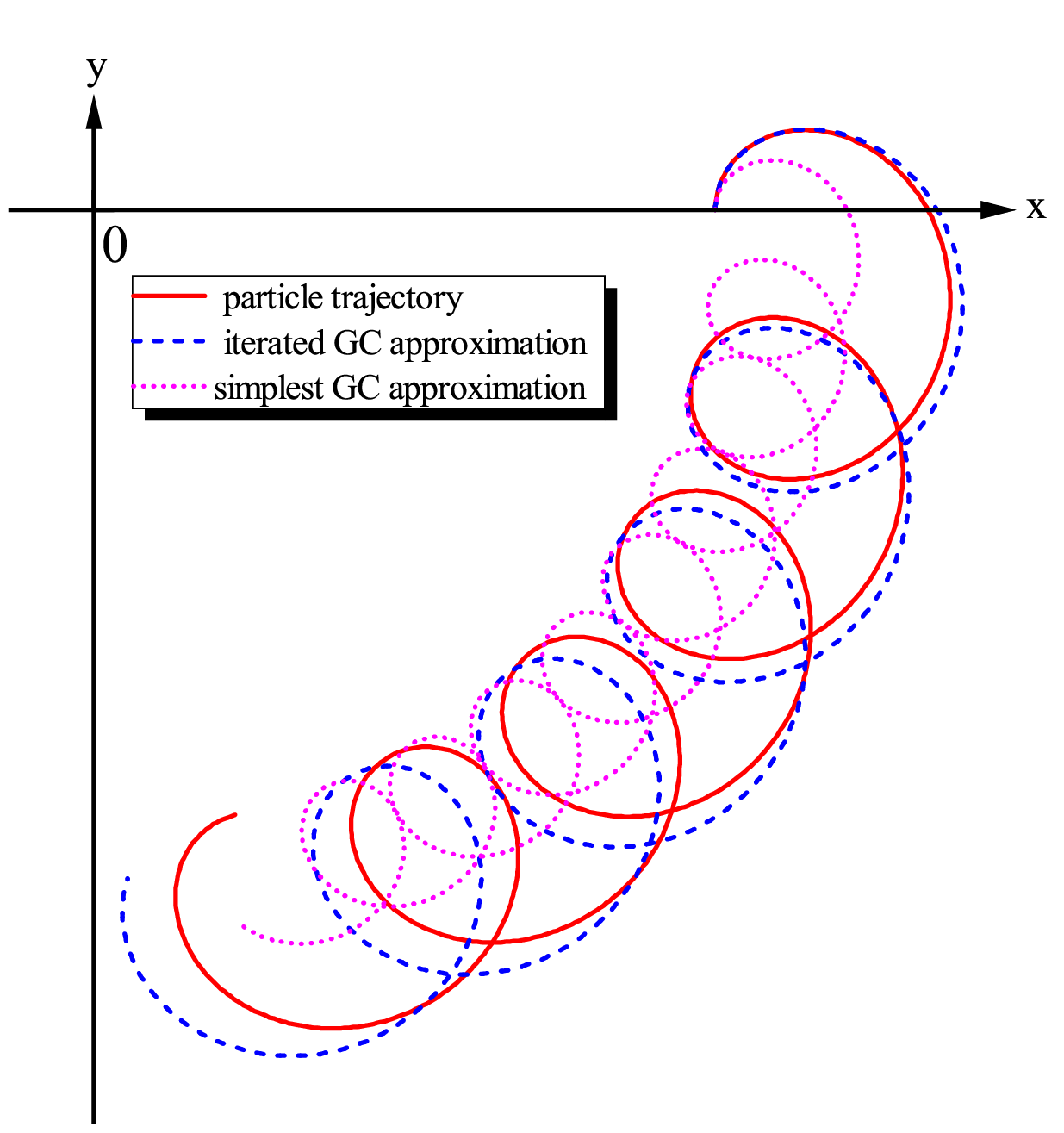}
  \caption{Trajectories computed with the simple GC approximation, with the iterated GC approximation and the exact trajectory for $v/v_*=0.15$. }
\label{comparison}
\end{figure}

Figure \ref{comparison} shows exact trajectory of the particle with red line, the simplest GC approximation ($i=0$ in equation (\ref{iter})) with pink dotted line, and GC approximation with converged parameters with blue dashed line for $v/v_*=0.15$. Although figure \ref{comparison} corresponds to a smaller speed than figure \ref{traj_def}, the approximation to the trajectory is not nearly as good as that in figure \ref{traj_def}. This is not surprising, considering that approximation (\ref{eqG1}) does not include any exact results. It is, however, obvious that iterating parameters for the GC approximation dramatically improved the results. It should also be noted that figure \ref{comparison} still corresponds to a very high speed which is 60\% of the confinement speed.
For smaller values of $v$, GC is considerably more accurate, as discussed further in  section \ref{accuracy}.
Because even the iterated frequencies still differ from the exact ones (\ref{gyro}) and (\ref{drift}), GC approximation suffers from secular error, and over many periods diverges from the exact trajectory. This type of error is also easily visible in figure \ref{comparison}.

\section{Accuracy of the epicycloid approximation}
\label{accuracy}

Next we analyze the accuracy of the epicycloid approximations introduced in the previous sections. First, we define the time-dependent distance $\delta$ between the exact trajectory and approximations described in sections \ref{sec_epi} and \ref{sec_GCa}:
\begin{equation}
\delta(t)=\sqrt{(x(t)-x_{exact}(t))^2+ (y(t)-y_{exact}(t))^2 } . \label{delta}
\end{equation}
Here $x(t)$ and $y(t)$ are given by equations  (\ref{eq1}) or  (\ref{eqG1}) and
the exact trajectory is obtained either analytically, as described in Ref. \onlinecite{kusaka,Avrett1962} or by a high accuracy numerical integration of the equations of motion.

\begin{figure}
 \includegraphics[scale=0.35]{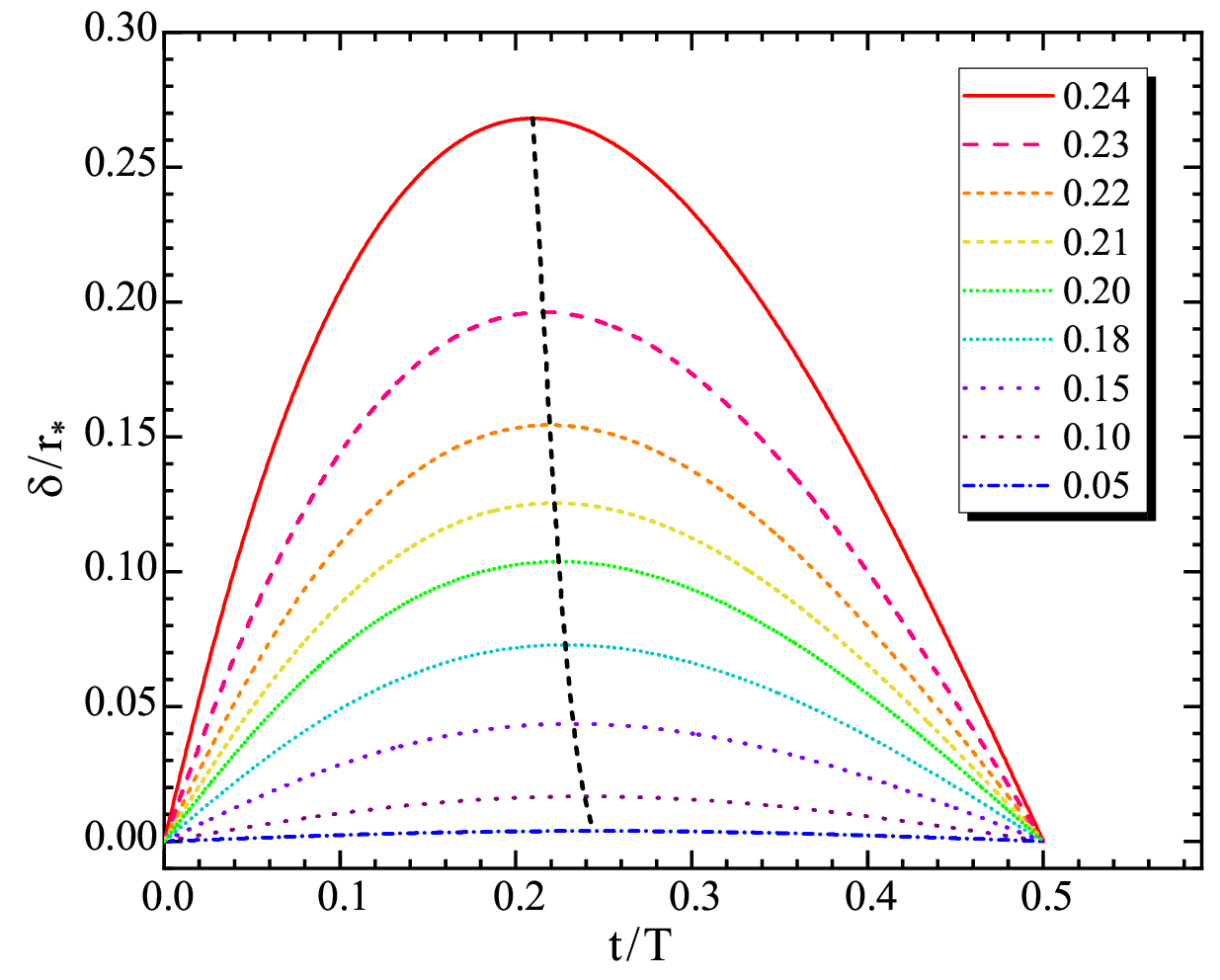}
  \caption{The approximation error for  (\ref{eq1})  as a function of time normalized with the gyration period, $T$.  The color code shows the values of $v/v_*$ for each curve. The black dashed line shows the location of the maximum error within the period.  }\label{del_T}
\end{figure}

Figure \ref{del_T} shows the approximation error $\delta$ of equation (\ref{eq1}) as a function of time for several values of $v/v_*$.  The time here is normalized with $T$ given by equation (\ref{gyro}), which is different for every value of the velocity. Because of symmetry, only half of the gyroperiod is shown.
As expected, the approximation error goes to zero at $r_{\min}$ and $r_{\max}$ and peaks roughly midway between these turning points.
The black dashed line in Figure  \ref{del_T} shows that the largest error, $\max (\delta)$, occurs at slightly different points during the cycle; for small $v$ the largest error occurs at quarter of the period. Comparing figures \ref{traj_def} and \ref{del_T} it is clear that the error shown in figure \ref{del_T} is larger than the distance between the exact and approximated trajectories in figure \ref{traj_def}.  This occurs because in equation (\ref{delta}) both the exact and
approximated position of the particle are used at the same moment in time;
thus $\delta$ is not exactly the geometrical distance between the two curves. If one is interested strictly in the distance between the two trajectories, without
regard for the time, then the appropriate estimation of accuracy would be just the geometrical distance between the two curves, calculated as
\begin{equation}
\delta_1(t)=\min_{\tau} \sqrt{(x(t)-x_{exact}(\tau))^2+ (y(t)-y_{exact}(\tau))^2 } . \label{delta1}
\end{equation}
Equation (\ref{delta1}) uses the numerically calculated trajectory as the reference curve and at its every point finds the smallest distance to the approximated trajectory.
It is obvious how this definition can be modified to use the approximated trajectory as the reference curve producing a similarly defined function $\delta_1 ' $.

\begin{figure}[tbph]
 \includegraphics[scale=0.35]{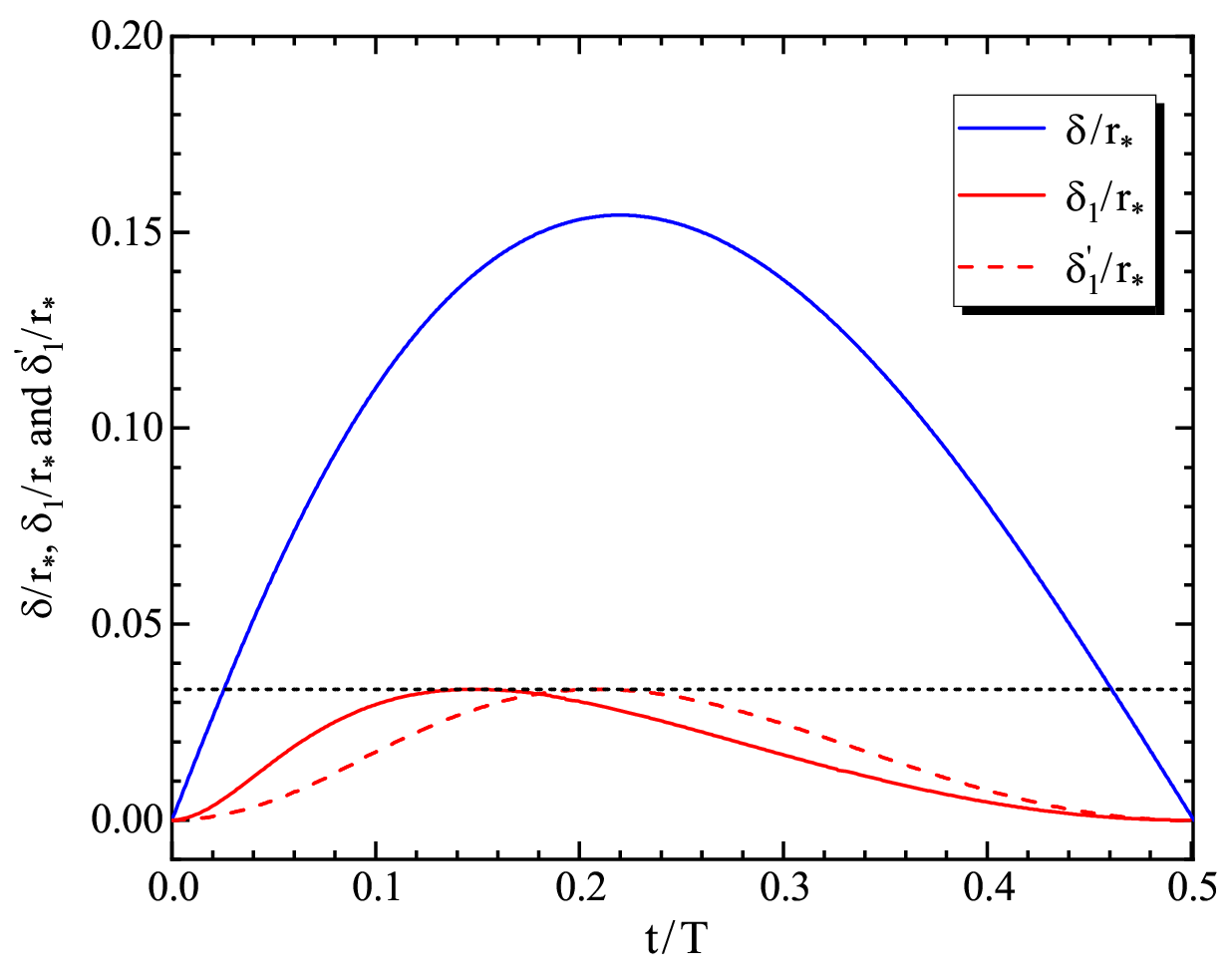}
  \caption{Comparison of $\delta$ (blue), $\delta_1$ (solid red line), and $\delta_1 ' $ (dashed red line) for $v=0.22 v_*$.  }\label{del_1}
\end{figure}

Figure \ref{del_1} shows the comparison between $\delta$ (blue line) defined by  (\ref{delta})  and the geometrical distance $\delta_1$ (red line) defined by  (\ref{delta1}).
For reference, $\delta_1 ' $ is also shown as a red dashed line; the maximum value of $\delta$ is, of course, the same as the maximum of $\delta_1 ' $, which is emphasized by a horizontal black dashed line. All plots are made for $v=0.22 v_*$.
Figure \ref{del_1} shows that for this speed $\delta_1$ is roughly a factor of 5 smaller than $\delta$ and peaks somewhat closer to $r_{\min}$. The value of $\delta_1$ is consistent with the distance
between the real and approximated trajectories seen in figure \ref{traj_def}. Thus, the accuracy of approximating the trajectory of the particle with epicycloid
(\ref{eq1}) is several times higher than approximating the position of the particle along the trajectory. The errors for the trajectory approximation (\ref{eqG1})
are qualitatively the same but larger.

\begin{figure}[tbph]
  \includegraphics[scale=0.35]{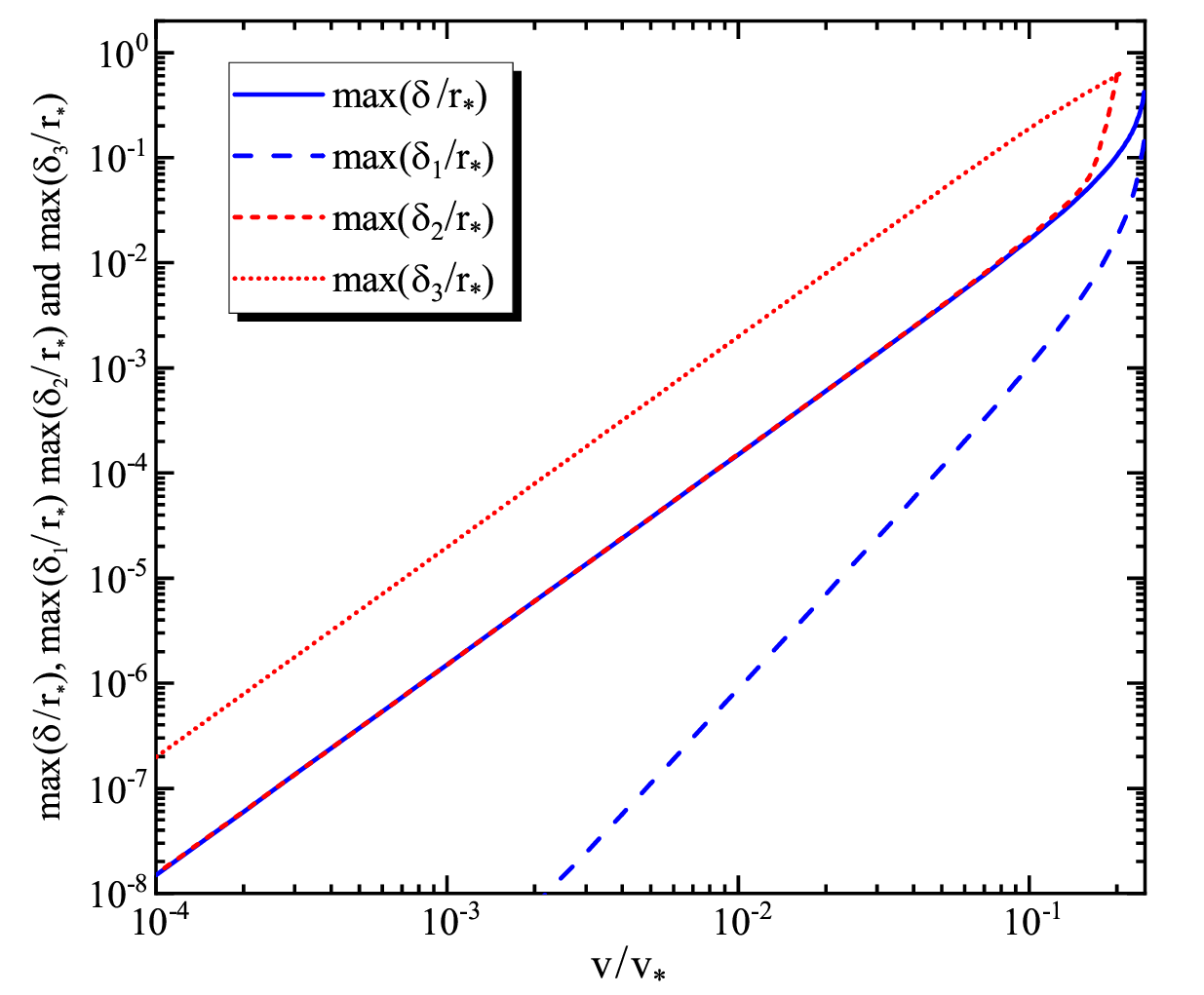}
  \caption{The accuracy of the approximation (\ref{eq1}) $\max(\delta$), shown with a solid blue line, and $\max(\delta_1)$, shown with a dashed blue line, as a function of $v/v_*$. Also shown the $\max(\delta$) measure of accuracy for the simple GC approximation (red dotted line) and the iterated GC approximation (red dashed line). }
\label{delta}
\end{figure}

It is natural to expect for approximations (\ref{eq1}) and  (\ref{eqG1}) to work best for small values of the particle velocity, when the GC approximation is most accurate, and to break down for large velocities approaching the confinement limit. This is illustrated by figure \ref{delta} which shows the $\max (\delta)$ with a solid blue line and $\max(\delta_1)$ with a dashed blue line as a function of $v/v_*$ for approximation (\ref{eq1}); note the logarithmic scale of this figure. Also shown are $\max (\delta_2)$
where $\delta_2$ is computed with equation (\ref{delta}) for
  the simple GC approximation, based on the instantaneous position of the particle (red dashed line), and $\max (\delta_3)$ for the iterated GC approximation (red dotted line).
For small $v$ the error $\max(\delta)\sim v^2$ for both approximations  (\ref{eqG1}) and (\ref{eq1}), while the geometric distance between the curves decreases at an even faster rate of $\max(\delta_1) \sim v^3$. Although the error for simplest GC approximation based on the particle position also decreases as $\sim v^2$, it is always roughly a factor of 10 larger than that for either the iterated GC or the epicycloid approximation. It is impressive, that for most of the considered range of $v$, the iterated GC approximation works nearly as well as the epicycloid approximation with the exact frequencies. The iterated GC approximation only loses it accuracy for high values of the velocities, and in fact  this approximation cannot be computed at all for $v> 0.204v_*$ when iterations (\ref{iter}) diverge. Figure \ref{delta} only shows the accuracy of all the approximations for the first gyro-cycle; because of the secular errors inherent in the GC approximation, over time it becomes progressively worse. The cycloid approximation with the exact frequencies is free of the secular errors. Furthermore, somewhat surprisingly for such a simple approximation, it remains reasonably accurate (on the order of 10\%) even at speeds close to the confinement threshold.

\section{Velocity}

Another important way in which the accuracy of approximations  (\ref{eq1}) and (\ref{eqG1}) can be quantified is by considering the velocity of the particle. Here we focus
on equation (\ref{eq1}), but the results for the GC approximation are quite similar.
During motion in magnetic field, in the absence of electric field, the velocity of the particle remains constant.
Differentiating (\ref{eq1}) we arrive at the following equation for the velocity for the epicycloid approximation:
\[v_{approx}^2=\left< v_{approx}^2 \right>+
\frac{r_{\max}^2-r_{\min}^2}{2} \Omega_D \omega \cos\left(\frac{2\pi t}{T}\right)  \]
where the average value of $v_{approx}^2$ over the period is given by
\begin{equation}
\left< v_{approx}^2 \right>= \left(\frac{r_{\max}+r_{\min}}{2}\Omega_D\right )^2 +\left(\frac{r_{\max}-r_{\min}}{2}\omega \right )^2. \label{v_ave}
\end{equation}

Unfortunately, $v_{approx}$ always depends on time, so approximation  (\ref{eq1})  is never exact. Expanding the expressions for $r_{\min}$, $r_{\max}$, etc. into Taylor series for small $v$ (see e.g. Ref. \onlinecite{Avrett1962}) it is easy to show that the second term in the above equation for $\left< v_{approx}^2 \right>$ dominates and this term approaches $v^2$ for small $v$.
This is not surprising, since the first term in  equation (\ref{v_ave}) represents the drift speed and the second the gyration speed. One of the requirements for GC approximation applicability is that the drift speed is much smaller than the particle speed \cite{northrop}.

\begin{figure}[tbph]
  \includegraphics[scale=0.35]{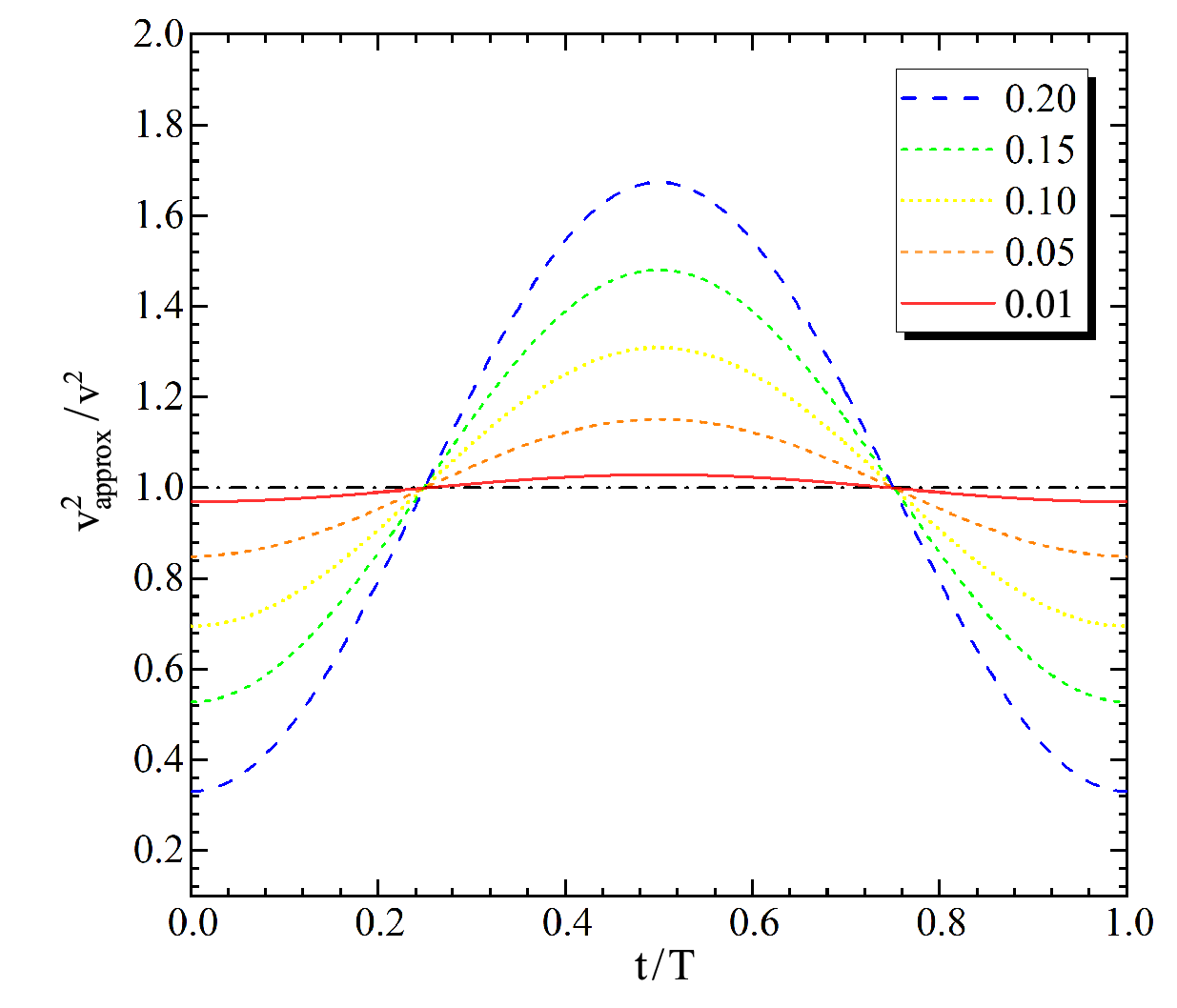}
  \caption{$v_{approx}^2/v^2$ as a function of time ($t/T$) for one period for several values of $v/v_*$ (color code).}
\label{v_app1}
\end{figure}

\begin{figure}[tbph]
  \includegraphics[scale=0.35]{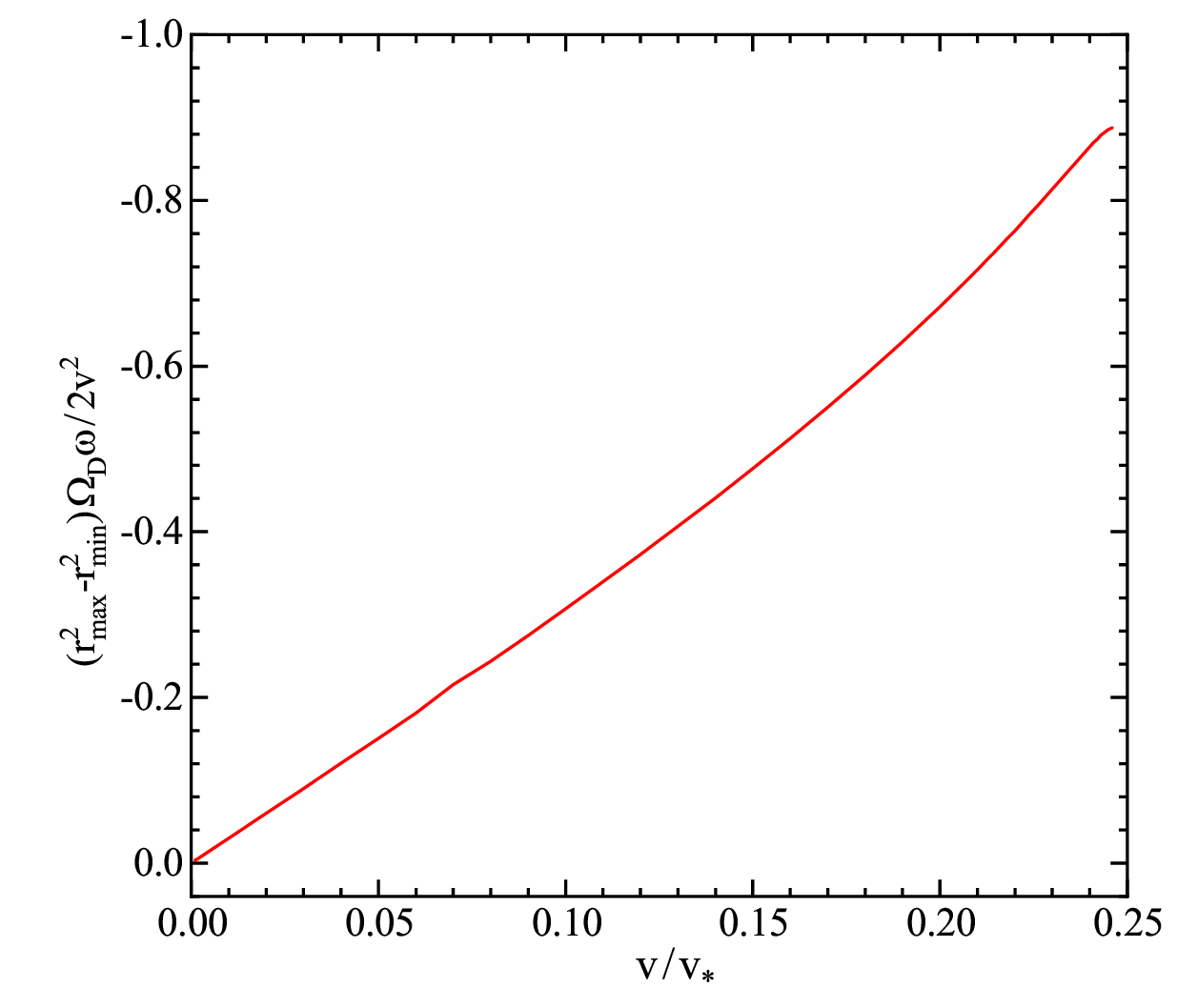}
  \caption{Normalized amplitude of the velocity oscillations as a function of $v/v_*$.}
\label{dV_T}
\end{figure}

Figure \ref{v_app1} shows $v_{approx}^2/v^2$ as a function of time ($t/T$) for one period for several values of $v/v_*$. As velocity of the particle deceases, the ratio $v_{approx}^2/v^2$ approaches horizontal line at value 1; however, for larger values of  $v/v_*$ the amplitude of the change in $v_{approx}$ is quite large.
The dependence of this amplitude  $(r_{\max}^2-r_{\min}^2) \Omega_D \omega / 2v^2$ on $v/v_*$ is shown in
figure \ref{dV_T}.  For small $v$ this normalized amplitude decreases linearly with $v$.

\begin{figure}[tbph]
  \includegraphics[scale=0.34]{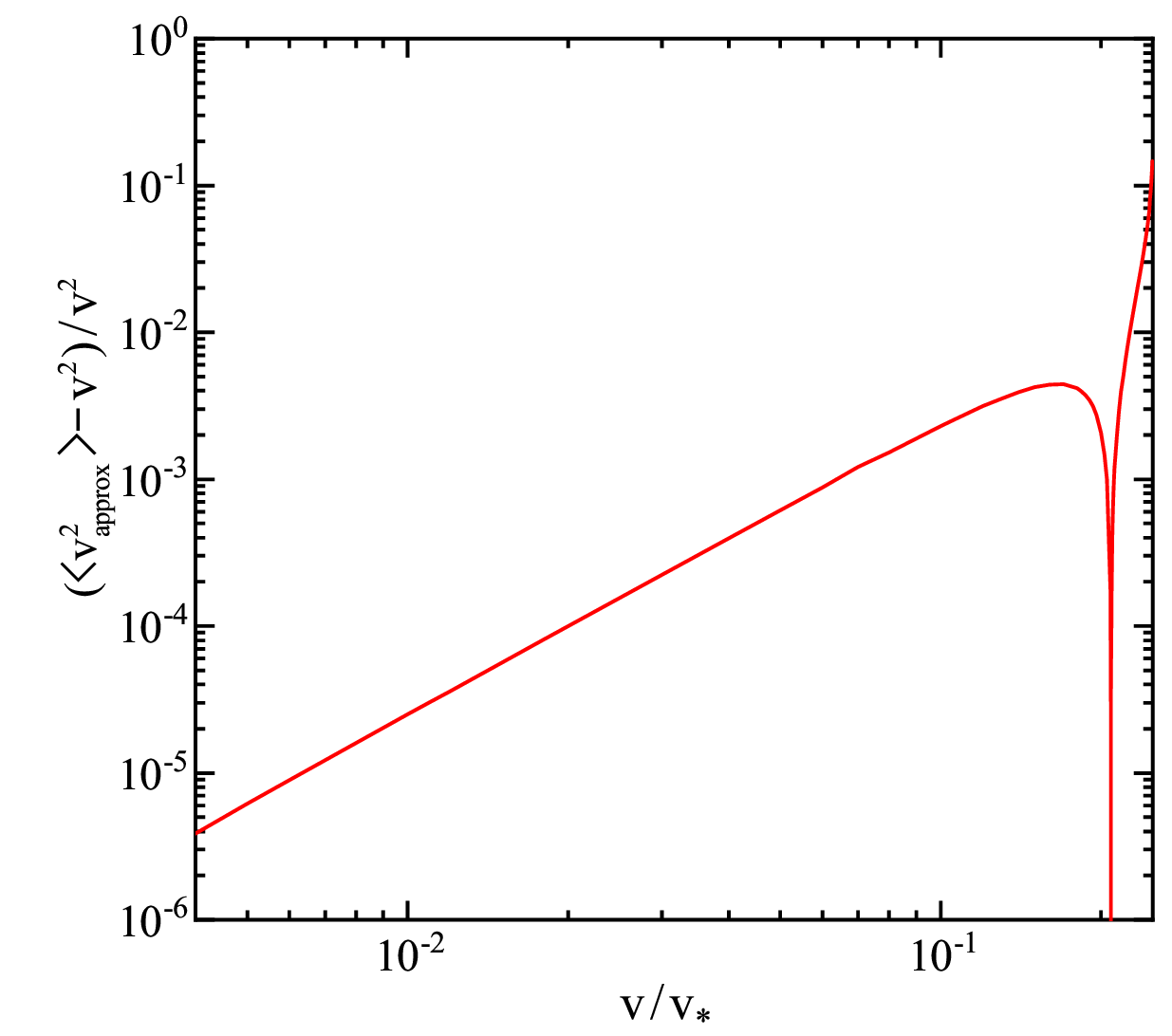}
  \caption{Error in the average velocity: $\left| \left< v_{approx}^2 \right>-v^2\right| /v^2$ as a function of $v/v_*$. }
\label{v_app2}
\end{figure}

Finally, figure \ref{v_app2}  illustrates the deviation of $\left< v_{approx}^2 \right>$ from $v^2$ as a function of $v/v_*$. The quantity
$\left| \left< v_{approx}^2 \right>-v^2\right| /v^2$ shown in figure \ref{v_app2} is the normalized deviation from the expected value.
For small values of $v$ this error decreases as $v^2$, the same rate as $\max(\delta)$ in figure \ref{delta}.

Comparison of figures \ref{dV_T} and \ref{v_app2} suggests that the largest cause of the error of approximation (\ref{eq1}) is likely associated with the amplitude of the velocity oscillations during the gyration. This conclusion is also supported be the fact that the trajectory itself is approximated by  (\ref{eq1}) better than the particle position along the trajectory, as indicated by figure \ref{delta}. It would be interesting to develop a modification to the epicycloid approximation which represents the velocity of the particle more accurately.

\section{Other axisymmetric magnetic fields}

\begin{figure}
  \includegraphics[scale=0.3]{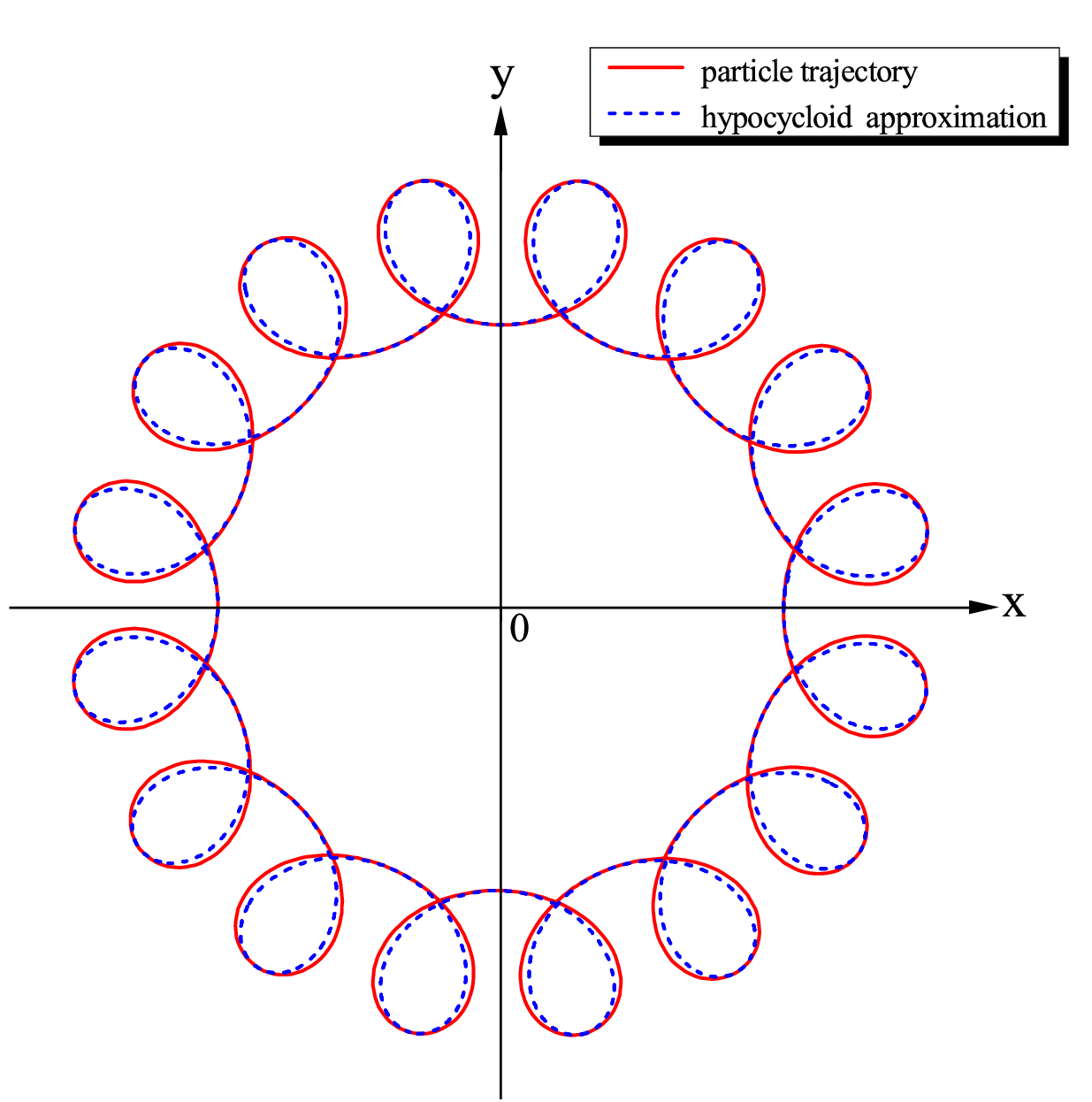}
 \includegraphics[scale=0.3]{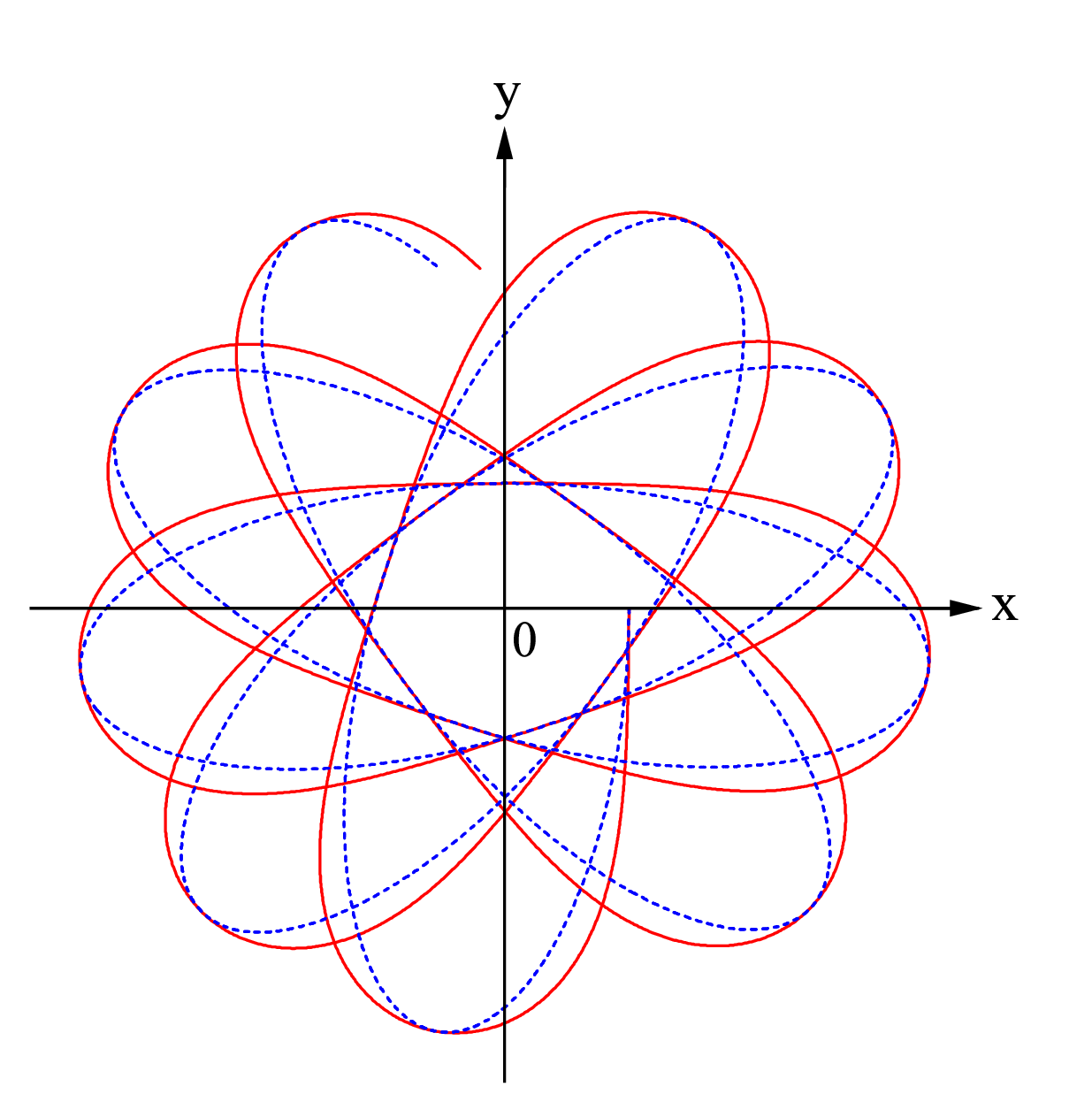}
 \caption{Comparison of the particle trajectory and the  hypocycloid  approximation   for $n=-3$, $A=10$ and $x_0=0.15$. Upper: $v_{y_0}=0.003$,  lower:
$v_{y_0}=-0.2$. Particle orbits on the lower panel enclose the origin, while those on the upper do not. }
\label{n=-3}
\end{figure}

Approximating particle trajectories with cycloidal curves can be applied to other axisymmetric fields as well. A simple generalization of (\ref{eq_B}) is
\begin{equation}
{B}_{z}=\frac{\cal M}{\left(x^{2}+y^{2}\right)^{n / 2}}. \label{B_n}
\end{equation}
Magnetic fields with such spatial dependencies were considered, for example, by  Ref. \onlinecite{Kabin2018}; $n=3$, obviously, corresponds to the equatorial plane of a magnetic dipole.
As long as $n>0$ the shape of the trajectories remains very similar to those discussed in this paper so far. The epicycloid approximations for these trajectories can be calculated in exactly the same way, with the only difference being that there are no known analytical solutions for the values of $n$ other than $3$ and $1$ \cite{kabin2015}. Therefore, the epicycloid parameters have to be calculated either from direct numerical integration of the particle motion for one period or from GC approximation.
The situation changes somewhat for $n<0$, when magnetic field intensity increases, rather than decreases, with distance. This behavior of the magnetic field changes the sign of the magnetic field gradient in (\ref{V_G}), and, therefore, the direction of the drift velocity.
Therefore, for $n<0$, the trajectory is represented by a hypocycloid \cite{Gordon1984}.
In fact, equations (\ref{par_abR}) still apply in these cases, however, parameter $b$ becomes negative in which case equation (\ref{eq1}) describes a  hypocycloid.
Figure \ref{n=-3} shows representative examples of such trajectories for $n=-3$. The curves in both upper and lower panes are hypocycloids, but the orbits in the  lower one enclose the origin, while the ones on the upper do not.
 The accuracy of the trajectory approximations remains qualitatively similar in all cases.

\section{Conclusions}

In this paper we described several approximations for trajectories of particles moving in the equatorial plane of a magnetic dipole. These approximations, inspired by the Guiding Center approach, are based on cycloids.
The best results are obtained, not surprisingly, when exact periods and gyration amplitudes are used. Approximation in which the gyroradius, gyrofrequency, and the drift period of a  particle are based on its initial position is by far the least accurate. Intermediate accuracy is achieved by using GC approximation in which the values of the gyroradius, gyrofrequency, and the drift period are iterated. For small values of the velocity, this approximation performs virtually as well as that based on the exact parameters, but for larger velocities it loses accuracy faster. Furthermore, since the radius of convergence of the gyrofrequency iterations is found to be smaller than the threshold velocity for particle confinement by the dipole field, this iteration procedure cannot be carried out for all values of initial velocity of the particle.
A detailed analysis of accuracy of these approximations shows that the trajectory of the particle is reproduced more accurately than its velocity. Although only motion in the equatorial field of a dipole is analyzed in detail, the results for other similar axisymmetric magnetic fields are virtually identical, except that epicycloids are replaced with hypocycloids in some cases.
In the future, it might be worthwhile to try to improve the representation of the velocity in cycloid-based approximations which would allow to design a more accurate yet still simple approximation.

\begin{acknowledgments}
This project is supported by the National Natural Science Foundation of
China (Grant Nos. 12065014 and 11705076), and by the HongLiu Support Funds for Excellent Youth
Talents of Lanzhou University of Technology.
\end{acknowledgments}

\nocite{*}

\begin{thebibliography}{}
%
%


 \bibitem{Lyons84}
 L. R. Lyons and  D. J.  Williams,
\emph{ Quantitative aspects of magnetospheric physics }
(D. Reidel, Dordrecht, Netherlands, 1984).


 \bibitem{Gombosi}
 T. I. Gombosi, 
\emph{Physics of the space environment }
(Cambridge University Press, Cambridge, 1998).


 \bibitem{northrop}
  T. G.  Northrop,
\emph{The adiabatic motion of charged particles }
(Interscience publishers, New York, 1963).


 \bibitem{li98}
X. Li,   D. N. Baker, M. Temerin,  G. D. Reeves and  R. D.  Belian,
"Simulation of dispersionless injections and drift echoes of energetic electrons associated with substorms,"
J. Geophys. Res.\textbf{25}, 
 3763-3766 (1998).

\bibitem{sarris2005}
T. Sarris and  X.  Li, 
"Evolution of the dispersionless injection boundary associated with substorms, "
Ann. Geophys. \textbf{23},
 877-884 
(2005).


 \bibitem{kabin2017}
K. Kabin,   G. Kalugin, E.  Spanswick  and  E. Donovan,
"Particle energization by a substorm dipolarization,"
J. Geophys. Res. \textbf{122},  349-368 (2017).

 \bibitem{kusaka}
C. Graef and  S. Kusaka, 
"On periodic orbits in the equatorial plane of a magnetic dipole "
J. Math. Phys. Massachusetts  \textbf{17},  43-54 (1938).


 \bibitem{Juarez1949}
 A. R. Juarez,
"Periods of motion in periodic orbits in the equatorial plane of a magnetic dipole,"
Phys. Rev.  \textbf{75(1)},  137-139  (1949).

\bibitem{Avrett1962}
 E. H. Avrett, 
"Particle motion in the equatorial plane of a dipole magnetic field, " 
J. Geophys. Res. \textbf{67},   53-58 (1962).


\bibitem{Kabin2019}
K. Kabin, 
"Two examples of exact calculations of the adiabatic invariant for charged particle motion in non-uniform axisymmetric magnetic fields, "
Physics of Plasmas   \textbf{26}, 
 012114 
(2019).


 \bibitem{johnston}
D. C.  Johnston, 
"Cycloidal paths in physics as superpositions of translational and rotational motions,"
Am. J. Phys   \textbf{87},  802-814 (2019).


 \bibitem{Kabin2018}
K. Kabin,   G. Kalugin, E.  Spanswick and E. Donovan,
"Threshold speed for two-dimensional confinement of charged particles in certain axisymmetric magnetic fields,"
Can. J. Phys. \textbf{96},
519-523
(2018).


\bibitem{burby2013}
J. W. Burby,   J. Squire, and  H. Qin, 
"Automation of the guiding center expansion," 
Phys. Plasmas 
 \textbf{20},  072105 (2013).




 \bibitem{kabin2015}
K. Kabin,  G.  Bonner, 
"Motion of a charged particle in an axisymmetric longitudinal magnetic field that is inversely proportional to the radius,"
Comp. Phys. Comm.  \textbf{189}, 155-161 (2015). 









\bibitem{Gordon1984}
 F. S. Gordon, S. P.  Gordon, 
"Mathematical Discovery via Computer Graphics: Hypocycloids and Epicycloids,"
The Two-Year College Mathematics Journal \textbf{15}, 
 440-443 
(1984).
https://www.tandfonline.com/doi/pdf/10.1080/00494925.1984.11972832





\end{thebibliography}


\end{document}